\documentclass[10pt,nofootinbib,showpacs,twocolumn,epsfig,aps,prd,amsmath,amssymb,superscriptaddress,showpacs]{revtex4-1}
\usepackage[utf8]{inputenc}

\usepackage{epsfig}
\usepackage{chngcntr}
\usepackage{wrapfig}

\usepackage{dcolumn} 
\usepackage{bm}            
\usepackage{subfigure}

\usepackage{graphicx}
\usepackage{longtable}
\usepackage{amsbsy,amsmath}
\usepackage{hyperref}
\usepackage{slashed}
\newcommand{\version}{\sc version 3.1.3,\ \ 10.12.2015.}

\renewcommand\Re{\operatorname{Re}}
\renewcommand\Im{\operatorname{Im}}

\begin{document}

\title{\Large Damping and Decoherence in Neutron Oscillations}

\thanks{\version} 

\author{B.O. Kerbikov} 
\email{borisk@itep.ru}

\author{M.S. Lukashov \medskip}
\email{m.s.lukashov@gmail.com}

\affiliation{{\sc A.I. Alikhanov Institute for Theoretical and Experimental Physics},\\ Bolshaya Cheremushhkinskaya 25, 117218, Moscow, Russia \medskip}

\affiliation{{\sc Moscow Institute of Physics and Technology},\\ Institutsky Pereulok 9, 141700, Dolgoprudny, Moscow Region, Russia \medskip}

\author{Y.A. Kamyshkov} 
\email{kamyshkov@utk.edu}

\author{L.J. Varriano \medskip}
\email{lvarrian@vols.utk.edu}

\affiliation{{\sc Department of Physics and Astronomy, University of Tennessee},\\ 1408 Circle Drive, Knoxville, TN 37996-1200, USA \medskip}

\date{\today}

\begin{abstract}
{ 
An analysis is made of the role played by the gas environment in neutron-mirror-neutron and neutron-antineutron oscillations. In the first process the interaction with the ambient medium induces a refraction energy shift which plays the role of an extra magnetic field. In the second process antineutron annihilation in practice might lead to strong decoherence, which should be taken into account in experiments with free neutrons looking for the neutron to antineutron transformation. 
}
%
%
\end{abstract}


\maketitle

\section{\label{sec:level1} introduction}

The observation of neutrons transforming into mirror neutrons ~\cite{Berezhiani:2005hv} or neutron transformation to antineutrons ~\cite{Kuzmin:1970} would be discoveries of fundamental importance. The idea that the Left-Right asymmetry of the Standard Model can be restored by the existence of the parallel mirror particle world has a long history starting from ~\cite{Lee:1956qn} and recounted in the recent review ~\cite{Okun:2006eb}. This concept bacame particularly important when it was realized that Mirror Matter can be a viable candidate for Dark Matter \cite{Berezhiani:2000gw}. Mirror and Ordinary Matter worlds have identical particle content and identical Standard Model self-interactions in their own sectors. Besides the evident gravitational force, the two sectors can also interact via neutral particle mixing such as neutron with mirror neutron, thus, leading to the $n \rightarrow n'$ oscillation phenomenon. That means that the wave function of the neutron particle has two components $n$ and $n'$, and the latter will develop with time under appropriate conditions such that after the ``measurement" of the particle, the neutron can be found to disapper  with some probability by converting into an undetectable sterile mirror particle. According to  ~\cite{Berezhiani:2005hv} such oscillations can occur rather quickly $\tau_{n,n'} \gtrsim 1$ s without contradiction with any particle or cosmological data.

The transformation of neutron to antineutron was extensively discussed in the recent paper ~\cite{Phillips:2015} in theoretical and experimental aspects. This process would violate baryon number B and $B-L$ by two units and might be key to understanding the matter-antimatter asymmetry in the universe. Neutron transformation to antineutron would demonstrate the Majorana nature of the neutron as it was originally conjectured in ~\cite{Majorana:1937}. 

Neutron transformation to mirror neutron could be observed as a process of neutron \textit{disappearance}. Neutron disappearance also can result from neutron swapping between our branewold and a neighboring one ~\cite{Sarrazin:2007}. On the other hand, the transformation of neutron to antineutron, using the language of neutrino physics, would be an \textit{appearance} process.

The aim of the present paper is to call attention to a physical process which, to our best knowledge, was not adequately taken into account in the analysis of the neutron oscillation experiments. This is the effect caused by collisions with the atoms or molecules of the residual gas inside the experimental setup. An important point in oscillation experiments is to provide conditions for degeneracy and coherence of the two states. To this end the magnetic field \textbf{B} is shielded to a minimum and the pressure inside the  storage chamber is minimized. It will be shown that collisions between neutrons and the residual gas molecules give rise to two effects. The first one is the energy splitting between the neutron and its partner. This may be described in terms of the refraction index and is equivalent to a complementary``magnetic field." The second one is the destruction of the off-diagonal coherence. This can make the interference between the two basic states impossible. 

We formulate the general formalism equally well suited for the description of both disappearance and appearance processes mentioned above. As an application, two examples are considered: (a) neutron-mirror-neutron oscillations neglecting the possible presence of mirror gas and mirror magnetic field and (b) a loss of coherence in neutron-antineutron oscillations.


\section{\label{sec:level2} Von Neumann-Liouville and Lindblad Equations}

The density matrix formalism is a natural way to describe the quantum system in contact with the environment. Interaction in the medium breaks the coherence of the propagation making the description in terms of the wave function impossible. Statistical matrix treatment has been used in a wide range of problems in atomic and particle physics. Our approach is close to that proposed in a seminal paper \cite{008} and later developed in \cite{009}. The problem of quantum damping or decoherence has been formulated in terms of the density matrix in \cite{010}. The application of the density matrix formalism to the description of matter effects in neutrino oscillations may be found in \cite{011}. 

In the Introduction, we have presented two posible neutron oscillation processes. To describe them, it is convenient to introduce the following notations. Let the oscillations occur between the two``flavor" states $\left|n_1\right>$ and $\left|n_2\right>$. By $\left|n_1\right>$ we mean the neutron, while $\left|n_2\right>$ may be mirror neutron, antineutron, or a neutron from another braneworld. There are also three possible scenarios for the interaction with the environment. Both states $\left|n_1\right>$ and $\left|n_2\right>$ may be embedded in the same medium. This is the case of neutron-antineutron oscillations in the presence of residual gas in a trap. The possible presence of a magnetic field will be different for neutron and antineutron 
as their magnetic moments are of opposite sign. For the neutron-mirror-neutron oscillations, the environments are different if we assume the presence of mirror gas and/or a mirror magnetic field. The same situation occurs for the neutron swapping between the two branes if the hidden brane is filled with another medium. Finally, only the state $\left|n_1\right>$ may be in contact with the environment like in neutron-mirror-neutron oscillations with mirror vacuum. This case, as we shall see from the equations, is described by the same formulas as the previous one with certain parameters set to zero. The formalism presented below will be applied to the situation when both basic states interact with the same environment. The situation when $\left|n_1\right>$ and $\left|n_2\right>$``feel" the influence of different reservoirs deserves a special investigation.

The two-state system is described by the density matrix \cite{012}.
\begin{equation}
\label{eq:01}
\hat \rho = 
\begin{pmatrix}
\varphi_1\varphi_1^* & \varphi_1\varphi_2^* \\
\varphi_1^*\varphi_2 & \varphi_2\varphi_2^* \\
\end{pmatrix}.
\end{equation}
Due to the interaction with the molecules of the reservoir and the decay of the basic states, the density matrix $\rho$ undergoes a non-unitary evolution. The \textit{von Neumann-Liouville equation} $\dot{\rho} = - i [H, \, \rho]$ is replaced by the \textit{Lindblad equation} \cite{013} for the reduced density matrix. It contains new terms due to the degrees of freedom of the environment which were integrated out. The properties of the environment are assumed to be unchanged by the interaction with the two-state system. Mathematical aspects of the problem \cite{013} will not be considered further here. Our aim is to determine how the presence of the surrounding gas changes the picture of the neutron oscillations. We shall follow the approach proposed in \cite{008}. The interaction of the $\left|n_1\right>$--$\left|n_2\right>$ system with the ambient gas molecules is described by the amplitude
\begin{equation}
\label{eq:02}
\hat{F}(\theta) = 
\begin{pmatrix}
f_1(\theta) & 0 \\
0 & f_2(\theta) \\
\end{pmatrix}.
\end{equation}
The reduced density matrix satisfies the following equation
\begin{equation}\label{eq:03}
\dfrac{d {\hat \rho}}{d t} = -i{\hat H}{\hat \rho} + i{\hat \rho}{\hat H}^{\dagger} + 2 \pi \nu v \int d (\cos\,\theta) {\hat F}(\theta) {\hat \rho} {\hat F}^{\dagger}(\theta).
\end{equation}
Here
\begin{widetext}
\begin{equation}\label{eq:04}
{\hat H} =
\begin{pmatrix}
E + \Delta_1 - \dfrac{2\pi}{k}\nu v f_1(\theta=0) - i\frac{\gamma}{2} & \varepsilon \\
\varepsilon & E + \Delta_2 - \dfrac{2\pi}{k}\nu v f_2(\theta=0) - i\frac{\gamma}{2} \\
\end{pmatrix}.
\end{equation}
\end{widetext}
In this Hamiltonian, $\Delta_i={\vec \mu_i}{\bf B}$ ($i=1,\, 2$) are the level shifts due to the magnetic field, $\gamma$ is the decay width of  $\left|n_1\right>$ and $\left|n_2\right>$  (assumed to be the same), $\nu = \dfrac{N}{V}$ is the gas number density. Here we consider that both $\left|n_1\right>$ and $\left|n_2\right>$  interact with the same ambient medium as it occurs in $n - {\bar n}$ oscillations. The diagonal elements of $\hat H$ contain the additional level shifts proportional to $f_i(\theta = 0)$ ($i=1,\, 2$) induced by the coherent forward scattering on the gas particles. When a neutron with momentum $k_0$ enters a medium, its momentum is modified according to $k'=n k_0$, where $n$ is the index of refraction. The expression for $n$ should incorporate the thermal motion of the gas particles. There is some discussion in the literature about how to take this into account \cite{014}. We follow the expression for $n$ presented in \cite{014} and \cite{015}
\begin{equation}\label{eq:05}
n = 1 + 2 \pi \nu \dfrac{m}{\mu}\dfrac{\left< f(k,\,\theta = 0) \right>}{k_0}.
\end{equation}
Here $m$ is the neutron mass, $\mu = m M (m + M)^{-1}$, $M$ is the mass of the target particle, $k = \mu v$, where $v = \left| {\bf v}_n - {\bf v}_t\right|$, and ${\bf v}_t$ is the velocity of the gas particle, $\left< \cdots \right>$ means the average over the velocity distribution of the gas particles. For the refraction level shift, one obtains
{\small \begin{equation}\label{eq:06}
\Delta E_i=\dfrac{1}{2m}\left[k_0^2 - (n k_0)^2\right] \simeq - \dfrac{2 \pi \nu}{\mu} {\left< f_i(k,\,\theta = 0) \right>},\,\, (i=1,\, 2).
\end{equation} }
This equation is an obvious generalization of the Fermi pseudopotential. The purpose of the present paper is to develop a general formalism. The application to specific experiments will be presented elsewhere. Therefore we take $v_t$ at a fixed value $v_t = \sqrt{\dfrac{3 T}{2 M}}$. For molecular hydrogen at room temperature $v_t \simeq 2.0 \cdot 10^5$ $cm \cdot s^{-1}$. Most of the experiments on neutron oscillations involve ultracold neutrons (UCN) with energy $E \lesssim 10^{-7}$ $eV$ and velocity $v_n \simeq (4-5) \cdot 10^2$ $cm \cdot s^{-1}$. Therefore $v = |{\bf v}_n - {\bf v}_t| \simeq v_t$ and one can replace $\left< f \right>$ by its value at $k = \mu v \simeq \mu v_t$. This is how to understand the expressions for the refraction level shifts in (\ref{eq:04}) written in the form proposed in \cite{008}. Next we notice that the $nH_2$ center-of-mass energy corresponding to $v \simeq v_t$ is only $E^{*}=\frac34\frac{k^2}{m} \simeq 0.014$ $eV$ and up to this energy the amplitude is almost momentum independent.

The collision integral in (\ref{eq:03}), the Lindblad term, is a price to pay for the use of the reduced density matrix describing only the evolution of the $\left|n_1\right> - \left|n_2\right>$ subsystem \cite{013}. Straightforward calculations lead to the following set of equations for the elements of the density matrix.
\begin{widetext}
{ \footnotesize \begin{align}
\dot{\rho}_{11} &=  -i\varepsilon(\rho_{21}-\rho_{12})-\gamma\rho_{11}, \label{eq:07} \\
\dot{\rho}_{12} &= -i\left[d - \dfrac{2\pi}{k}\nu v \Re(f_1-f_2)-4\pi \nu v \Im f_1f_2^*\right]\rho_{12} - 4 \pi \nu v \left[\Im\left(\dfrac{f_1+f_2}{2k}\right)-\Re f_1^*f_2\right]\rho_{12} -i\varepsilon(\rho_{22}-\rho_{11})-\gamma\rho_{12}, \label{eq:08} \\
\dot{\rho}_{21} &= -i\left[d - \dfrac{2\pi}{k}\nu v \Re(f_1-f_2)-4\pi \nu v \Im f_1f_2^*\right]\rho_{21} - 4 \pi \nu v \left[\Im\left(\dfrac{f_1+f_2}{2k}\right)-\Re f_1^*f_2\right]\rho_{21} + i\varepsilon(\rho_{22}-\rho_{11})-\gamma\rho_{21}, \label{eq:09} \\
\dot{\rho}_{22} &= +i\varepsilon(\rho_{21}-\rho_{12})-\gamma\rho_{22}. \label{eq:10}
\end{align} }
\end{widetext}
where $d = \Delta_1 - \Delta_2$. A closer look at Eqs. (\ref{eq:07}-\ref{eq:10}) allows one to trace the origin of the coefficients governing the time evolution of the coherence matrix elements $\rho_{12}$ and $\rho_{21}$. Consider the following quantity 

\begin{equation}\label{eq:11}
{\rm \Lambda} = i \nu v \dfrac{\pi}{k^2} (1 - S_1^*S_2),
\end{equation}
where $S_i$ ($i=1,\, 2$) are the scattering matrices in the channels $\left|n_1\right>$ and $\left|n_2\right>$. Then 
\begin{equation}\label{eq:12}
\Re \Lambda = -4 \pi \nu v \left[\Re\left(\dfrac{f_1-f_2}{2k}\right)+\Im f_1f_2^*\right],
\end{equation}
\begin{equation}\label{eq:13}
\Im \Lambda = +4 \pi \nu v \left[\Im\left(\dfrac{f_1+f_2}{2k}\right)-\Re f_1f_2^*\right].
\end{equation}

\section{\label{sec:level3} Bloch Equations}

The time evolution of the density matrix given by (\ref{eq:03}) and (\ref{eq:07}-\ref{eq:10}) may be represented in a vector form of the Bloch equation \cite{010, 016}. The real Bloch $3$-vector $\mathbf{R}$ is introduced by the expansion of the density matrix over the Pauli matrices
\begin{equation}\label{eq:14}
\rho = \dfrac12 (\hat{1} + \mathbf R{\vec\sigma}),
\end{equation}
where
\begin{equation}\label{eq:15}
\mathbf R =
\begin{pmatrix}
\rho_{12} + \rho_{21} \\
-i(\rho_{21} - \rho_{12}) \\
\rho_{11} - \rho_{22} \\
\end{pmatrix}.
\end{equation}

The set of the equations of motion (\ref{eq:07}-\ref{eq:10}) is equivalent to the following equation of motion for the Bloch vector
\begin{equation}\label{eq:16}
\dot{\mathbf R} = {\mathbf V} \times {\mathbf R} - {\mathrm D}_T {\mathbf R}_T - \gamma  {\mathbf R},
\end{equation}
where
{ \footnotesize
\begin{equation}\label{eq:17}
{\mathbf V} =
\begin{pmatrix}
2 \epsilon \\
0 \\
d + \Re \Lambda \\
\end{pmatrix},\quad 
{\mathrm D}_T = 
\begin{pmatrix}
\Im \Lambda & 0 \\
0 & \Im \Lambda \\
\end{pmatrix},\quad 
 {\mathbf R}_T = 
\begin{pmatrix}
R_x \\
R_y \\
\end{pmatrix}.
\end{equation} }
The first term in (\ref{eq:16}) describes the ``precession" of the ``polarization vector" $\mathbf R$ around the ``magnetic field" $\mathbf V$, the second term corresponds to the transverse damping, and the last one gives the exponential decay of the basic states. Due the second and the third terms, the Bloch vector $\mathbf R$ shrinks in length. The damping parameter ${\mathrm D}_T$ characterizes the rate at which the classical environment destroys the off-diagonal elements of $\rho$, thus leading to a loss of coherence.

From the structure of Eqs. (\ref{eq:16})-(\ref{eq:17}), it follows that the quantity $\Re \Lambda$ may be viewed as a complementary ``magnetic field," which is added to the magnetic field driven term $d$. In other words, one can argue that $\Re \Lambda$ corresponds to the energy shift due the refraction index. The damping parameter ${\mathrm D}_T$, which is proportional to $\Im \Lambda$, induces decoherence, thus suppressing the oscillations. On the other hand, $\Im \Lambda$ may be understood in terms of continuous ``measurements," i.e., the transmission of the information to the gas environment. 

\section{\label{sec:level4} Application to Experiments}

As noted in the Introduction, the application of the present formalism to the experiments will be the subject of another publication. Here, we will set the scene for this task. The searches for neutron-mirror-neutron oscillations have been performed in \cite{017}-\cite{019}. The review of the past and future experiments on nucleon-antinucleon oscillations may be found in \cite{Phillips:2015}. We shall estimate the relevant parameters entering into  (\ref{eq:07} - \ref{eq:10}) and  (\ref{eq:16} - \ref{eq:17}) for the case of UCN confined in a trap {\footnote{See \cite{020} for a comparison of the storage and beam experiments.}}. The favorable condition for oscillations is the degeneracy of the two levels. In order to reach this regime, the magnetic field $\textbf{B}$ is shielded to a minimum (``zero'' field run), and the pressure inside the storage chamber is minimized. 


Let us estimate the effect of the residual gas in the experiments  \cite{017, 018} on the search for neutron-mirror-neutron oscillations. Let the residual gas be molecular hydrogen at room temperature with average thermal velocity $v_t \simeq 2 \cdot 10^5$ $\frac{cm}{s}$, which, within our approximation, is equal to $nH_2$ relative velocity. In the experiment \cite{017} the pressure was $P_1 \simeq 10^{-8}$ $atm$, in \cite{018} $P_2 \simeq 10^{-6}$ $atm$. Respectively, at room temperature, the densities were $\nu_1 \simeq 2.5 \cdot 10^{11}$ $cm^{-3}$ and $\nu_2 \simeq 2.5 \cdot 10^{13}$ $cm^{-3}$. Molecular hydrogen is a mixture of ortho- and para-hydrogen. The internuclear distance is $R \simeq 0.74 \cdot 10^{-8}$ $cm$, the moment of inertia is $I \simeq (0.015\,\,eV)^{-1}$. At $T >> I^{-1}$ one has $\frac{\nu\,(para)}{\nu\,{ortho}} = \frac{1}{3}$. At room temperature, this ratio is ${2.91}^{-1}$ \cite{021}. The requisite energy to excite the molecule from the ground para-hydrogen rotational state to the ground ortho-hydrogen one is $E^{*}_{min}=I^{-1}$ which is slightly higher than the $nH_2$ c. m. energy  $E^{*}=\dfrac{3\,k^2}{4\,m}\simeq0.014\,eV$. We assume that the hydrogen gas is in a normal state when all molecules are in their ground state and para-to-ortho fraction is $\frac{1}{3}$. Corrections to this approximation will be considered in another publication. 

The coherent scattering amplitude for the normal mixture is \cite{022,023,024} \footnote[2] {Note that the signs of the amplitude and the scattering length are opposite in \cite{026}, which is not the case in the definition of the amplitude in \cite{023}}.
\begin{equation}\label{eq:18}
f = -2\,\left(\dfrac{3}{4}a_t + \dfrac{1}{4}a_s\right) \simeq 3.74\,fm \,\,
\end{equation}
The scattering from the two protons forming the $H_2$ molecule adds coherently if the de Broglie wavelength in the c. m. system $\lambda = \frac{2\pi}{k} \simeq \frac{3\pi}{m\,v_t}\simeq 3.0 \cdot 10^{-8}$ $cm$ is much larger than the internuclear distance $R$. In the conditions under consideration $\frac{\lambda}{R} \simeq 4$. Another important parameter is the ratio of $\lambda$ to the mean distance $\nu^{-\frac{1}{3}}$ between the nearest scattering centers, $\Lambda = \dfrac{\lambda}{\nu^{-\frac{1}{3}}}$ (de Boer delocalization length). In both experiments \cite{017, 018} $\Lambda << 1$.

We take the value (\ref{eq:18}) for the amplitude $f_1$ and neglect the possible presence of the mirror gas \cite{Foot:2002} and hence $f_2=0$. Then
\begin{equation}\label{eq:19}
\Re \Lambda = - 4 \pi \nu_i v \Re\left(\frac{f_1}{2k}\right)\simeq 
\begin{cases}
-0.4 \cdot 10^{-18}\,\,eV,&\text{when i=1},\\
-0.4 \cdot 10^{-16}\,\,eV,&\text{when i=2},\\
\end{cases}
\end{equation}
for the experiments \cite{017} and \cite{018} respectively. According to Eqs. (\ref{eq:08}-\ref{eq:09}) and (\ref{eq:17}), these parameters have to be compared with $d = \Delta_1 = |\mu_n B|$ at the minimum value of $B_0$ reached in the experiments (the presence of mirror magnetic field is neglected). In \cite{017, 018} $B_0 \simeq 10\,\,nT$, i.e., 
\begin{equation}\label{eq:20}
|\mu_n B_0| = 6 \cdot 10^{-12} \left( \frac{B}{1 G} \right)eV= 0.6 \cdot 10^{-15}\,eV.
\end{equation}
Therefore the ``effective magnetic field'' (see (\ref{eq:17})) induced by the residual gas introduces $\sim10\,\%$ correction to the value of $B_0$ in \cite{018}. Note also that $\Re \Lambda \gg \frac{1}{\tau_{nn'}}$, where $\tau_{nn'}^{-1}$ is the $n-n'$ mixing parameter, $\tau_{nn'}^{-1} < 1.5 \cdot 10^{-18}\,eV$ \cite{017}.

From (\ref{eq:16}-\ref{eq:17}) it follows that when $\Im \Lambda = 0$, the residual gas induces only an additional level splitting but does not lead to decoherence. In terms of the density matrix, decoherence destroys the off-diagonal terms and hence the interference between the basic states. This is explicitly seen from (\ref{eq:08}) and (\ref{eq:09}). The terms $\rho_{12}$ and $\rho_{21}$ are damped by $\Im \Lambda$ and by the $\beta$-decay constant $\gamma$. Decoherence is important in $n - {\bar n}$ oscillations due to the strong annihilation in the residual gas. The leading term in $\Im \Lambda$ (see (\ref{eq:13})) is 
\begin{equation}\label{eq:21}
\Im \Lambda \simeq 4 \pi \nu v \Im (\frac{f_2}{2k}) = \frac{1}{2}\nu v \sigma_a,
\end{equation}
where $\sigma_a$ is the annihilation cross section. At low energy according to PDG $v \sigma_a ({\bar n}p) \simeq (50-55)$ $mb$. Using this value and taking $\nu_{2} = 2.5 \cdot 10^{13}$ $cm^{-3}$ \cite{018} and $\nu_{3} =5 \cdot 10^{10}$  $cm^{-3}$ \cite{027},  one obtains 
\begin{equation}\label{eq:22}
(\Im \Lambda)_{2} \simeq 10^{-2}\,s^{-1}  \simeq 10^{-17}\,eV.
\end{equation}
\begin{equation}\label{eq:23}
(\Im \Lambda)_{3} \simeq 10^{-5}\,s^{-1}  \simeq 10^{-20}\,eV.
\end{equation}

This is an important result. It means that
\begin{equation}\label{eq:24}
\Im \Lambda \gg  \frac{1}{\tau_{n{\bar n}}} = \varepsilon \lesssim 10^{-23}\,eV.
\end{equation}

This inequality  means that the decoherence is so strong that it may damp $n-{\bar n}$ oscillations. This problem will be  thoughtfully investigated in the work in progress. Here, we present an oversimplified but very transparent picture of damped oscillations. Let us for simplicity assume that magnetic field, i.e., the term $d$ in (\ref{eq:17}) is exactly zero, and more than  that, $\Re \Lambda = 0$. Then Eq. (\ref{eq:16}) for $R_z = \rho_{11} - \rho_{22}$ can be easily recast into the following second order one
\begin{equation}\label{eq:25}
\dfrac{d^2}{dt^2}R_z + (2 \gamma+\lambda) \dfrac{d}{dt}R_z + (\gamma^{2}+\gamma \lambda +4\varepsilon^2) R_z = 0,
\end{equation}
where $\lambda \equiv \Im \Lambda$. This is an equation for the oscillator with friction. Assuming that at $t=0$ the system is in the state $\left| n_1 \right>$, so that $\rho_{11}=1$, $\rho_{22}=0$, the solution of (\ref{eq:25}) is
\begin{equation}\label{eq:26}
R_z(t)\,=\,e^{-\gamma t} \,e^{-\frac{1}{2}\lambda t} \left( \cosh \Omega t + \dfrac{\lambda}{4\,\Omega}\,\sinh \Omega t \right),
 \end{equation}
where
\begin{equation}\label{eq:27}
\Omega^2 = \dfrac14\,\lambda^2 - 4\,\varepsilon^2 > 0.
\end{equation}

In the limit $\lambda \gg \varepsilon$ (see  (\ref{eq:22}),  (\ref{eq:23}),  (\ref{eq:24})) the overdamping solution of  (\ref{eq:25}) has the form

{\large \begin{equation}\label{eq:28}
R_z \sim e^{-\gamma t} \cdot e^{-\dfrac{4\varepsilon^2}{\lambda}t}.
\end{equation} }

Solution (\ref{eq:28}) corresponds to a regime without oscillations. The physical reason is that the time scale for annihilation is much shorter than for oscillations. Vector $R_z$ does not have enough time to turn from $R_z = 1$, $\rho_{11}=1$, $\rho_{22}=0$ to $R_z = -1$, $\rho_{11}=0$, $\rho_{22}=1$ \cite{010}. To overcome overdamping, the pressure of the residual gas must be several orders of magnitude lower than the value e.g. $P = 10^{-6}$ $atm$ used above. As will be shown elsewhere, in its main features this picture is correct without the simplifying assumptions which have been used. Note that the limit $\varepsilon \lesssim 10^{-23}$ $eV$ was obtained not only in the ILL experiments \cite{027} in which the pressure was $P \simeq 2.0 \cdot 10^{-9}$ $atm$ and the damping could wash out the oscillations, but also from the stability of nuclei \cite{SuperK:2015}. In $n \rightarrow \bar{n}$ search experiments like \cite{027} or the proposed new ESS-based experiment \cite{Phillips:2015} with free neutron cold beams, the neutron observation time is a fraction of a second. It might be important to account for the characteristic damping time  $\frac {1}{Im \Lambda}$ of the order of $10^{2} - 10^{5}$ sec like in (\ref{eq:22}), (\ref{eq:23}) together with other factors that we have so far neglected for simplicity. Detailed calculations of these effects will be described in a future work in progress. 

\section{\label{sec:level5} Conclusion and Prospects}

In this paper, we have developed a general approach to neutron oscillations in a gas environment. At the core of our formalism are von Neumann-Liouville and Bloch equations, which correctly describe the two-level system embedded in a gas reservoir. The interaction with the environment results in two effects. First, a refraction level shift is induced which plays the role of an extra magnetic field. Neutron-mirror-neutron oscillation is a process where this phenomenon might be important. Second, the coherence, and hence oscillations, may be completely destroyed in the regime of overdamping. This is the case for neutron-antineutron oscillations unless the residual gas pressure is extremely low. The practical implications of this effect for the future ESS-based experiment \cite{Phillips:2015} will be discussed in detail in the work in progress. Finally, we mention that the problem of neutron interaction with the surrounding surfaces \cite{028} was also left for future, more-detailed elaboration.

\begin{acknowledgments}
We thank Z. Berezhiani, A. Dolgov, D. Gorbunov, A. Serebrov, and A. Voronin for valuable discussions and remarks. We are grateful to I. Khriplovich for his unpublished note on damping. The work is supported in part by Russian Foundation for Basic Research Grant $14-02-00395$ and US DOE Grant DE-SC0014558. 
\end{acknowledgments}



\begin{thebibliography}{99}

\bibitem{Berezhiani:2005hv} 
  Z.~Berezhiani and L.~Bento,
  Phys.\ Rev.\ Lett.\  {\bf 96}, 081801 (2006).\smallskip

\bibitem{Kuzmin:1970} V.A.~Kuzmin, JETP Lett. \textbf{12}, 228 (1970). \smallskip

\bibitem{Lee:1956qn} 
  T.D.~Lee and C.N.~Yang,
  Phys.\ Rev.\  {\textbf 104}, 254 (1956). \smallskip

\bibitem{Okun:2006eb} L.B. Okun, Phys. Usp. \textbf{50}, 380 (2007).\smallskip

\bibitem{Berezhiani:2000gw}
  Z.~Berezhiani, D.~Comelli and F.~L.~Villante,
  Phys.\ Lett.\ B {\textbf 503}, 362 (2001). \smallskip

\bibitem{Phillips:2015}
 D.~G.~Phillips, II {\textit et al.},
 Submitted to: Phys.Rept., arXiv:1410.1100 [hep-ex]. \smallskip

\bibitem{Majorana:1937}
  E.~Majorana,
  Nuovo Cim.\  {\textbf 14}, 171 (1937).

\bibitem{Sarrazin:2007} M. Sarrazin  F. Petit, Int. J. Mod. Phys. \textbf{A22}, 2629 (2007); \\  M. Sarrazin et al., Phys. Rev. \textbf{D 91}, 075013 (2015).\smallskip

\bibitem{008} G. Feinberg and S. Weinberg, Phys. Rev. \textbf{123}, 1439 (1961).\smallskip

\bibitem{009} S.V. Demidov, D.S. Gorbunov, A.A. Tokareva, Phys. Rev. \textbf{D 85}, 015022 (2012).\smallskip

\bibitem{010} L. Stodolsky, \textit{``Quantum Damping and its Paradoxes''} in \textit{``Quantum Coherence''}, ed. J.S.Anandan (World Scientific, Singapore, 1990).\smallskip

\bibitem{011} A. Dolgov, Sov. J. Nucl. Phys. \textbf{33}, 700 (1981); \\
L. Stodolsky, Phys. Rev. \textbf{D 36}, 2273 (1987); \\
A.Yu. Smirnov, Phys. Scripta \textbf{T 121}, 57 (2005).\smallskip

\bibitem{012} R.P. Feynman, \textit{Statistical Mechanics: A Set of Lectures} (W.A. Benjamin Inc., Mass., 1972).\smallskip

\bibitem{013} A. Kossakowski, Rep. Math. Phys. \textbf{3}, 247 (1972); \\ G. Lindblad, Commun. Math. Phys. \textbf{48}, 119 (1976).\smallskip

\bibitem{014} C. Champenois et al., Phys. Rev. \textbf{A 77}, 013621 (2008).\smallskip

\bibitem{015} B. Vacchini and K. Hornberger, Physics Reports, \textbf{478}, 71 (2009).\smallskip

\bibitem{016} F. Bloch, Phys. Rev. \textbf{70}, 460 (1946); \\
R. Feynman, F. Vernon, R. Hellwarth, Jour. of Appl. Phys. \textbf{28}, 49 (1957).\smallskip

\bibitem{017} A.P. Serebrov et al., Phys. Lett. \textbf{B 663}, 181 (2008).\smallskip 

\bibitem{018} G. Ban et al., Phys. Rev. Lett. \textbf{99}, 161603 (2007).\smallskip 

\bibitem{019} I. Altarev et al., Phys. Rev. \textbf{D 80}, 032003 (2009).\smallskip 

\bibitem{020} Yu. N. Pokotilovski, Phys. Lett.\textbf{B 693}, 214 (2006).\smallskip 

\bibitem{021} J.A. Young and J.U. Koppel, Phys. Rev. \textbf{135}, A 603 (1964).\smallskip 

\bibitem{022} J. Schwinger and E. Teller, Phys. Rev. \textbf{52}, 286 (1937).\smallskip 

\bibitem{023} G.L. Squires and A.T. Stewart, Proc. Roy. Soc. \textbf{230}, 19 (1955).\smallskip 

\bibitem{024} V.F. Sears, \textit{Neutron News}, \textbf{3}, 26 (1992).\smallskip 

\bibitem{026} L.D. Landau \& E.M. Lifshitz, \textit{Quantum Mechanics: Non-Relativistic Theory} (Pergamon Press, London, 1977).\smallskip 

\bibitem{Foot:2002} R. Foot, \textit{``Does Mirror Matter Exist?''}, [arXiv:hep-ph/0207175].\smallskip

\bibitem{027} M. Baldo-Ceolin et al., Z. Phys. \textbf{C 63}, 409 (1994).\smallskip 

\bibitem{SuperK:2015}
 K.~Abe {\it et al.} [Super-Kamiokande Collaboration],
  Phys.\ Rev.\ D {\bf 91}, 072006 (2015).\smallskip 

\bibitem{028} B. Kerbikov, Phys. At. Nucl. \textbf{66}, 2178 (2003); \\
B.O. Kerbikov, A.E. Kudryavtsev and V.A. Lensky, J. Exp. Theor. Phys. \textbf{98}, 417 (2004); \\ B. Kerbikov and O. Lychkovskiy, Phys. Rev. \textbf{C 77}, 065504 (2008).\smallskip 





\end{thebibliography}

\end{document}